\documentclass{article}

\usepackage[numbers]{natbib}
\usepackage{graphicx}
\usepackage{float}
\usepackage{multirow}
\usepackage[most]{tcolorbox}
\usepackage{caption}
\newcounter{mybox}

\usepackage[dblblindworkshop,final]{neurips_2025}

\usepackage[utf8]{inputenc} 
\usepackage[T1]{fontenc}    
\usepackage{hyperref}       
\usepackage{url}            
\usepackage{booktabs}       
\usepackage{amsfonts}       
\usepackage{nicefrac}       
\usepackage{microtype}      
\usepackage{xcolor}         

\workshoptitle{The Second Workshop on GenAI for Health: Potential, Trust, and Policy Compliance}

\title{Can You Spot the Virtual Patient (VP)? Expert Evaluation, Turing Test, Linguistic Analysis, and Semantic Similarity Analysis}

\author{%
  Reyhaneh Hosseinpourkhoshkbari\\
  School of Behavioral and Brain Sciences\\
  University of Texas at Dallas \\
  \texttt{reyhaneh.hosseinpour@utdallas.edu} \\
  \And
   Wei-chen Huang\\
 School of Behavioral and Brain Sciences\\
  University of Texas at Dallas \\
  \texttt{Wei-Chen.Huang@UTDallas.edu} \\
  \And
   Suvel Muttreja\\
  School of Behavioral and Brain Sciences\\
  University of Texas at Dallas \\
  \texttt{sxm200218@utdallas.edu} \\
    \And
   Richard M.Golden\\
  School of Behavioral and Brain Sciences \\
  University of Texas at Dallas \\
  \texttt{golden@utdallas.edu} 
}

\begin{document}
\begingroup
\renewcommand\thefootnote{}\footnote{A shorter, non-archival version of this paper was accepted as a poster to The AHLI Machine Learning for Health (ML4H) Symposium 2025 \citep{hosseinpour2025virtualpatient}.}%
\addtocounter{footnote}{-1}%
\endgroup

\maketitle

\vspace{-0.5cm}

\begin{abstract}
  Communication is a core clinical competency and essential for safe, effective patient care. Virtual patients (VPs) powered by large language models (LLMs) offer a promising alternative to traditional formats such as standardized patients (SPs). However, few studies have quantitatively as well as qualitatively evaluated their realism across diverse clinical scenarios using behavioral studies and linguistic dialog analyses. This study evaluated the realism of GPT-4o-generated VPs through expert review, Turing testing, linguistic analysis, and semantic analysis. We generated 44 VPs using structured illness scripts derived from doctor–simulated patient dialogues across 17 clinical categories. Experts annotated responses for hallucinations, omissions, and repetitions, with high interrater reliability ($ICC > 0.77$).A Turing test revealed that participants struggled to distinguish VP responses from those of real patients, with average classification accuracy for VP responses falling below chance level.  Lexical and syntactic analyses of over 2,000 conversation turns showed that VP responses were generally realistic—more formal and lexically consistent—while human responses displayed greater emotional variation. Semantic similarity using BioClinicalBERT and cosine similarity averaged 0.871 (response level) and 0.842 (transcript level).  Our findings support the integration of LLM-based VPs in communication training while highlighting areas for trust, transparency, and refinement—contributing to both GenAI deployment in healthcare and its safety evaluation.
\end{abstract}

\vspace{-0.5cm}
\section{Introduction}
\vspace{-0.25cm}

Teaching and assessment of clinical communication skills are core components of medical school curricula, as doctor-patient interactions lie at the heart of medical practice \citep{hosseinpourkhoshkbari2025clinical}. \citet{kurtz2003marrying} emphasizes that communication skills should be trained alongside other consultation skills, such as clinical reasoning, history taking, and clinical examination, to enhance teaching effectiveness and ensure practical application in medical settings.
Recently, simulated environments have become an essential component in medical education for training and evaluating clinical skills. In these settings, medical students interact with a "simulated patient", which may be a human actor, a digital learning tool, or an artificial intelligence (AI) system, to demonstrate and refine their skills \cite{ten2018approaches}.
Standardized patients (SPs) are human-simulated patients who are trained to portray particular medical conditions for educational objectives \citet{barrows1993overview}. SPs play a crucial role in medical education, and their impact on student outcomes may provide long-term benefits \cite{johnson2020impact}; however, establishing and sustaining a high-quality SP program can be both expensive and time-consuming \cite{mavis2006developments}. Virtual patients (VPs) are a type of simulated patient that can be incorporated into interactive computer programs. Such computer programs can be, in turn, used to simulate a variety of real-life clinical scenarios. VPs allow any clinical learner to interact with an agent at any time of day and at any location. They can also be integrated into advanced software platforms that provide automated learner assessment and feedback. By offering a safe environment, VP systems help learners to improve their clinical abilities without risking patient safety.

In this paper, we present a comprehensive framework for evaluating the realism and reliability of VPs generated by GPT-4o for use in clinical communication training. Our approach consists of four core components:
\begin{itemize}
    \item \textbf{Structured VP Generation.} We develop a structured prompting strategy grounded in illness scripts extracted from simulated doctor--human-patient interactions. This enables the generation of 44 diverse VPs spanning 17 clinical conditions, offering a scalable and reproducible method for creating synthetic clinical dialogue agents.
    \item \textbf{Expert-Based Quality Evaluation.} To evaluate the quality of VP responses, we define five expert-reviewed metrics: omissions (missing expected information), inappropriate repetitions (redundant or unnecessary content), hallucinations (factually incorrect statements), successful turns (accurate and contextually appropriate responses), and total turns. Two clinical educators annotated 1,094 conversation turns independently following an initial calibration round. We report strong interrater agreement using intraclass correlation coefficients (ICC), supporting the robustness of these quality metrics.
    \item \textbf{Behavioral Turing Test.} We conduct a behavioral evaluation based on a Turing Test framework \cite{turing1950computing}, examining whether human raters can distinguish between real and VP-generated patient responses. Inspired by prior work on LLM-human indistinguishability \cite{wu2024self,jones2024people,jones2025large,rathi2024gpt}, this study also investigates the impact of providing raters with diagnostic hints, simulating real-world conditions where clinicians may use heuristics to assess patient authenticity.
    \item \textbf{Linguistic and Syntactic Analysis.} We conduct a detailed linguistic analysis comparing VP and human responses across lexical and syntactic dimensions. Although LLMs are explicitly trained to generate human-like language, few studies have rigorously examined how closely their outputs resemble real patient language in clinical contexts \cite{martinez2024beware,park2025does,munoz2024contrasting}. We compute a range of lexical diversity metrics to assess vocabulary variation and analyze part-of-speech distributions to evaluate syntactic structure.
    \item \textbf{Semantic Similarity.} To assess semantic alignment between VP and human responses, we computed cosine similarity scores using contextual embeddings generated by BioClinicalBERT \citep{alsentzer2019publicly}. This domain-specific, representation-based approach allows for automated evaluation of dialogue content similarity without requiring manual annotation or human judgment \citep{reimers2019sentence}.
\end{itemize}

\vspace{-0.5cm}
\section{Related work}
\vspace{-0.25cm}

Recent advancements in LLMs have facilitated the development of VPs in medical education.
 For example, \citet{holderried2024generative} used structured prompts based on illness scripts covering five categories and assessed realism through script consistency, medical plausibility, and student feedback. They found that ChatGPT can simulate a VP experience with mostly plausible responses and a generally positive user experience, though occasional implausible information was noted. \citet{aster2025chatgpt} used short, unstructured prompts that were symptom-focused. The realism of their developed VP was evaluated using a short post-interaction questionnaire measuring students’ perceived autonomy during history taking (e.g., freedom of choice and task relevance) and their prior experience using ChatGPT. \citet{borg2024enhancing} applied a detailed, structured prompt, including patient description and prior dialog turns; while students preferred this format over non-interactive alternatives, delays and repetitive responses reduced perceived realism. \citet{yi2025feasibility} used prompts with limited patient detail and relied on expert ratings for response quality across five dimensions (relevance, accuracy, fluency, succinctness, and impersonation), reporting some plausible answers. \citet{yuan2025ai} introduced a VP system but did not specify the prompt design used to generate responses. Their evaluation focused on usability before and after optimization, showing improvements in personality and user experience, yet clinical appropriateness and accuracy across diverse populations were not assessed.
 While these studies report promising results in learner engagement and VP realism, several challenges remain. One challenge is the variability in prompt design across studies, which ranges from brief clinical scenarios to highly detailed ones, and few offer guidance on optimal prompt design for consistency and realism \cite{holderried2024generative,li2024leveraging, aster2025chatgpt}. Consequently, response delays caused by lengthy prompts can disrupt the natural flow of VP interactions \cite{irfan2025between, borg2024creating, grevisse2024raspatient}. In addition, the VP often breaks character and reverts to its default assistant role, especially when students fail to provide clear clinical cues \cite{grevisse2024raspatient}. “Hallucinations” (i.e., generating confusing or meaningless responses) continue to pose a threat to learning outcomes \cite{ji2023survey}. Furthermore, existing evaluations primarily rely on human ratings of empathy, response appropriateness, and consistency with predefined patient profiles. However, they are often limited to a single clinical case and a few conversational turns, overlooking long-term educational outcomes \cite{holderried2024generative,li2024leveraging,aster2025chatgpt}. 

\vspace{-0.25cm}
\section{Methodology}

\vspace{-0.25cm}
\begin{figure}[!ht]
  \centering
  \includegraphics[width=0.8\textwidth]{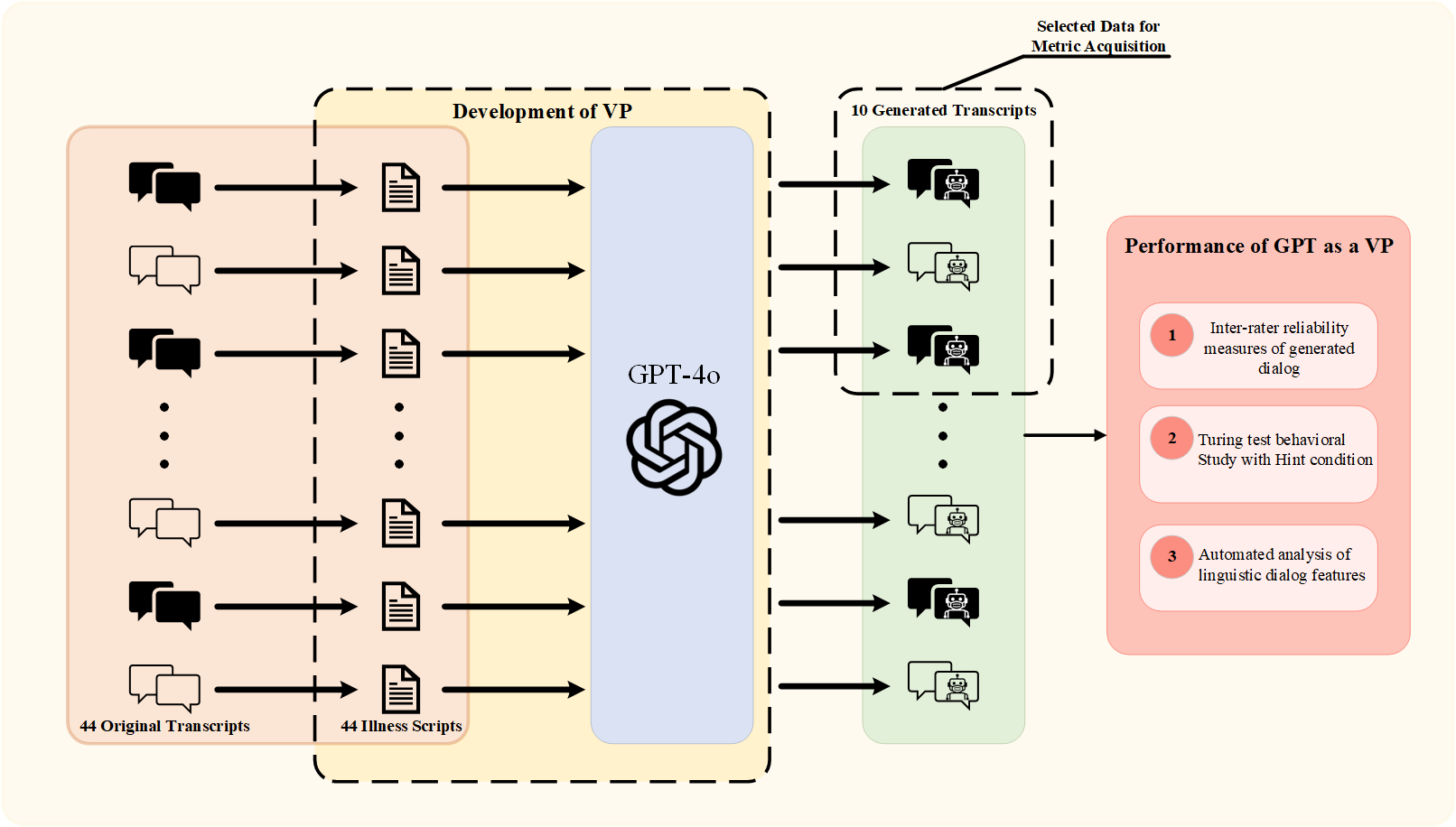} 
  \caption{Overview of the VP development and evaluation pipeline.}
  \label{fig:Diagram1}
\end{figure}

\subsection{Materials}
This section outlines the materials, procedures, and evaluation methods employed in the development and assessment of VP dialogues. An overview of the methodological framework is presented in Figure \ref{fig:Diagram1}.
\subsubsection{Dataset}
The \citet{fareez2022dataset} dataset, which was used to develop the illness scripts, consists of transcripts from simulated conversations between senior Canadian medical students (acting as "doctors") and resident doctors (acting as "patients"). While the dataset includes a variety of medical cases, it is predominantly focused on respiratory cases, which make up 78.7\% of the transcripts. The remaining cases cover musculoskeletal (16.9\%), gastrointestinal (2.2\%), cardiac (1.8\%), and dermatological (0.4\%) issues. For this study, a subset of 44 transcripts specifically related to respiratory cases was used. These transcripts consisted of a total of 2,139 conversation turns (question–answer pairs).

\subsubsection{ Large language model}
In this study, we employed zero-shot prompting, a strategy where the model is given a well-designed prompt that clearly defines the task without including any specific examples \citep{radford2019language}. We developed a series of illness scripts and utilized them to prompt ChatGPT-4o to generate diverse VP dialogs based on the dataset provided by \citet{fareez2022dataset}. The VP were generated using OpenAI's ChatGPT-4o via the API, with default parameters (temperature = 1.0, top-p = 1.0, presence penalty = 0, frequency penalty = 0).
\subsubsection{Development of the VP }
The illness script for each patient was created by completing a structured template that included key details about the patient’s main symptoms, medical history, social history, and so on (see an example of illness script in \nameref{appendix:appendixa}, Box \ref{box:example_script}). To enhance the quality and flow of the dialogue, additional lines were added before and after the core patient details, providing context and creating more natural interactions. The prompts for ChatGPT were created by completing an illness script template using data from transcripts of 44 respiratory cases, ensuring realistic and clinically grounded simulations. Some transcripts lack key patient details such as name, age, or occupation. To maintain a smooth VP interaction and prevent it from breaking character or reverting to its default ChatGPT persona \cite{grevisse2024raspatient}, we label these fields as “unknown to the transcript." By guiding the VP to “make up” a relevant answer when transcript data is missing, we preserve realism and instructional continuity during simulated clinical encounters. To initiate the conversation with the VP, we use the same doctor utterances from the original clinical transcripts \cite{fareez2022dataset}. (Refer to \nameref{appendix:appendixa} for illustrative examples of doctor–human patient versus doctor–VP conversations.)

\subsection{Performance of GPT as a virtual patient}
To assess whether the quality of GPT's responses was suitable for a VP experience, we employed three evaluation methods.
\subsubsection{Human rater classification of VP response types}
The quality of responses generated by the VP was evaluated using a structured set of metrics developed by two expert reviewers. 
First, both coders independently reviewed and coded an initial set of 10 transcripts, which included interactions between the VP and the doctor. After completing their independent coding, the two coders compared their annotations to identify discrepancies and refine their understanding of the metrics. These metrics were derived from an initial analysis of 10 VP transcripts and included: (1) the number of turns with omissions (missing expected information), (2) the number of turns with inappropriate repetitions (redundant or unnecessary content), (3) the number of turns with hallucinations (false, fabricated, or misleading information) (4) the number of successfully completed turns (accurate and contextually appropriate responses), and (5) the total number of conversational turns (see \nameref{appendix:appendixc} for example of common hallucinations, repetitions, and omissions). Moreover, metrics like "missing expected information, inappropriate repetitions, and hallucinations have been recognized as a concern in the previous studies \cite{sardesai2024utilizing,holderried2024generative, holderried2024language, li2024leveraging, grevisse2024raspatient, yi2025feasibility}.  
 Once a coding system was developed by the two coders, independent coding was carried out on the remaining 21 VP transcripts (1094 conversation turns). The intraclass correlation coefficient (ICC) was used to evaluate the consistency of numerical ratings between two raters. 

\subsubsection{Turing test}

 In recent years, several studies have used the Turing Test framework to assess the conversational abilities of LLMs, examining how well they can generate human-like dialogue \cite{turing1950computing, wu2024self,jones2024people, jones2025large,rathi2024gpt,uchendu2021turingbench}.
Building on this approach, we developed an experimental survey consisting of 20 dialog-based items: 10 drawn from real human–patient interactions and 10 generated by a VP in response to the same physician inputs. This balanced mix allowed for an assessment of whether participants could accurately distinguish between human-simulated patient and VP dialogs. 

The protocol for this study was reviewed and approved by the Institutional Review Board at the University of Texas at Dallas (IRB-24-871). All participants provided informed consent prior to participation. The data used were fully de-identified.

The study involved 50 undergraduate psychology students (N = 50), all enrolled in an English-medium institution. Sample size was determined by a power analysis assuming an effect size of 0.70, power of 0.80, and $\alpha$ = 0.05 \cite{creswell2015educational}. Before beginning the main task, participants completed a short tutorial with practice items. To assess the influence of guidance, we employed a between-subjects design: 25 participants received a brief written hint to assist in classification (see Box~\ref{box:Hint}), while the remaining 25 received no such aid. The hint was derived from 1,045 annotated conversation turns and informed by prior analysis \cite{sardesai2024utilizing}. Including this condition enabled us to evaluate the sensitivity of the experimental task; if participants performed better with the hint, this would support the task’s ability to detect subtle differences between human and VP dialogues.

The experiment also incorporated within-subject measurements. For each trial, they rated their confidence on a 4-point Likert scale (1 = not confident, 4 = very confident), and reaction times were logged. These measures allowed us to evaluate not only classification accuracy but also participants’ metacognitive assessments and decision-making behavior.

\begin{tcolorbox}[
  colback=yellow!10,
  colframe=orange!80!black,
  title=Box \refstepcounter{mybox}\label{box:Hint}\themybox: \textbf{HINT provided to human participants},
  fonttitle=\bfseries,
  boxrule=0.65pt,
  sharp corners=south,
  before skip=10pt, after skip=10pt,
  left=6pt, right=6pt, top=6pt, bottom=6pt
]
Keep in mind that computer-generated responses tend to be more formal and structured than human responses. For example, the computer will tend to avoid filler words such as “Um” and “Ah”. The computer also will tend to avoid repeating words. Additionally, computer responses may sometimes be a bit longer and more detailed compared to those from humans.
\end{tcolorbox}

\subsubsection{Linguistic features of conversations}

To assess the trustworthiness and linguistic authenticity of VP outputs, we performed a fine-grained lexical and grammatical analysis, comparing distributions of syntactic categories and vocabulary richness with human data. This Lexical diversity was assessed using multiple metrics from the LexicalRichness Python library \cite{lex}, including Type-Token Ratio (TTR) \cite{malvern2004lexical}, Root TTR (RTTR), and the Maas index \cite{van2007comparing,tweedie1998variable}. While TTR measures the ratio of unique words to total words, it is sensitive to text length, especially in short or highly variable responses \cite{munoz2024contrasting}. RTTR and Maas mitigate this issue by normalizing for response length.

To further reduce sensitivity to length and capture deeper vocabulary variation, we included the Measure of Textual Lexical Diversity (MTLD) and Hypergeometric Distribution Diversity (HDD) \cite{mccarthy2010mtld}, both of which are designed to handle variable-length texts more robustly. We also computed Yule’s K, which quantifies lexical concentration and is unaffected by text length, and the Moving Average TTR (MATTR), which analyzes diversity over sliding word windows \cite{covington2010cutting}.

These metrics were applied to both full transcripts ($n = 44$) and individual utterances ($n = 2{,}194$), allowing for lexical comparisons at multiple levels of granularity. Additionally, we performed part-of-speech (POS) tagging to examine syntactic composition \cite{cutting1992practical,divyapushpalakshmi2021efficient,kumawat2015pos}. POS tags—such as nouns, verbs, adjectives, and adverbs—were analyzed to compare the grammatical distributions between human and VP responses, offering complementary insights into their linguistic profiles.

\subsubsection{Semantic analysis}

To quantitatively assess the semantic alignment between virtual patient (VP) and real human responses, we conducted a domain-specific similarity analysis using \textbf{BioClinicalBERT} \citep{alsentzer2019publicly}, a transformer-based language model pre-trained on clinical notes and biomedical literature. This model was selected over general-purpose encoders due to its superior ability to capture medically relevant concepts and discourse patterns.

We then computed pairwise cosine similarity between these embeddings to evaluate semantic proximity \citep{reimers2019sentence}. This approach enables fully automated, representation-based comparison of clinical dialogues without requiring manual annotation, offering a scalable and reproducible method for benchmarking VP realism.

\vspace{-0.25cm}
\section{Results}
\vspace{-0.25cm}

In this section, we present the results of our comprehensive evaluation of the proposed VP, focusing on expert assessment, Turing test performance, and key linguistic characteristics of VP responses.
\subsection{Human rater classification of VP response types}
As shown in Table~\ref{tab:ICC}, we observed at least \textit{good} interrater reliability across all three error types: hallucinations, omissions, and repetitions. Intraclass correlation coefficient (ICC) analyses indicated \textit{excellent} agreement for hallucination ratings and \textit{good to excellent} agreement for omissions and repetitions, confirming consistency in expert judgments.
Across all 21 VP transcripts, the average rate of successfully completed conversation turns was 96.6\%, with individual transcripts ranging from 90\% to 100\%. This high completion rate reflects the overall relevance and coherence of the VP responses. Among the three error types, hallucinations were the most frequent, occurring in an average of 2.8\% of turns. In contrast, omissions and repetitions were rare, with both metrics exhibiting median values of 0\% and mean rates of 0.31\% and 0.26\%, respectively. Although hallucinations were the most common error, their low absolute frequency suggests that the VP responses were generally accurate and appropriate. Overall, the low incidence of omissions and repetitions further reinforces the quality and reliability of the generated interactions.
\begin{table}[h!]
\centering
\renewcommand{\arraystretch}{1.3}
\begin{tabular}{@{}llccl@{}}
\toprule
\textbf{Metrics} & \textbf{ICC} & \textbf{95\% CI}  & \textbf{p-value} & \textbf{Interpretation} \\ \midrule
 Hallucination & 0.814 & [0.603 , 0.920] & $p < .0001 (***)$ & Excellent reliability \\
Omissions                 & 0.783 & [0.544 , 0.905] & $p < .0001 (***)$ & Good--Excellent \\
Repetition                 & 0.774 & [0.528 , 0.901] & $p < .0001 (***)$ & Good reliability \\
\bottomrule
\end{tabular}
\caption{Summary of inter-rater reliability for three variables using ICC.}
\label{tab:ICC}
\end{table}

\subsection{Turing test results}
A two-way ANOVA was conducted to examine the effects of \textit{hint} (with vs.\ without) and \textit{dialog type} (human vs.\ VP) on participants’ accuracy in identifying the source of a dialog. Participants correctly classified human-simulated patient dialogs (M = 81.2\%) significantly more often than VP dialogs (M = 42\%), $F(1, 90) = 8.69$, $p = .004$. Those who received a hint performed better overall (M = 68\%) than those who did not (M = 55\%), $F(1, 90) = 77.34$, $p < .001$. However, there was no significant interaction between hint condition and dialog type, $F(1, 90) = 2.05$, $p = .156$.

\begin{figure}[!ht]
  \centering
  \includegraphics[width=0.73\textwidth]{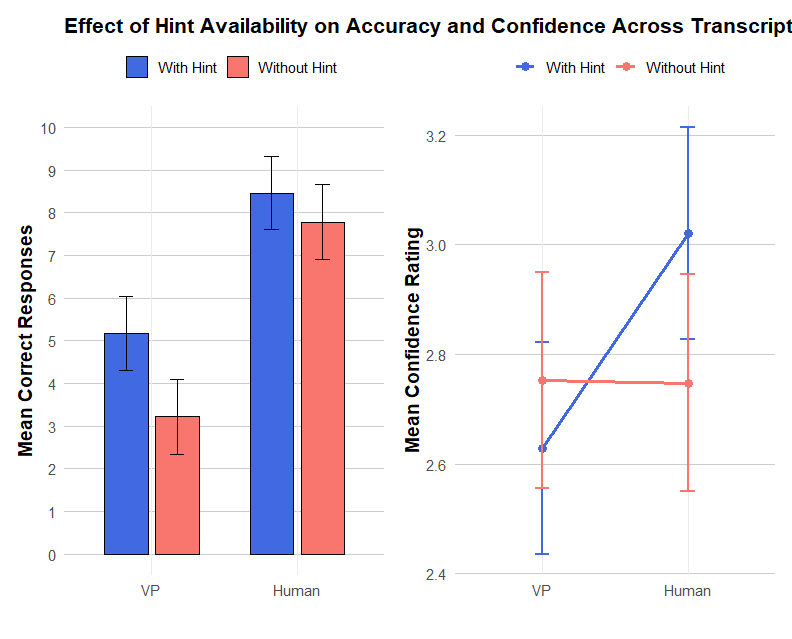} 
  \caption{Comparison of participants’ accuracy and confidence across VP and human transcripts, with and without hint conditions.}
  \label{fig:two_plots}
\end{figure}

As shown in Figure~\ref{fig:two_plots} (left), participants provided with a hint were more accurate in classifying both VP responses (M = 5.17, SE = 0.44, CI [4.29, 6.04]) and human dialogs (M = 8.46, SE = 0.44, CI [7.58, 9.33]) than those without a hint (VP: M = 3.22, SE = 0.45, CI [2.32, 4.11]; human: M = 7.78, SE = 0.45, CI [6.89, 8.68]). Pairwise comparisons revealed that both groups struggled more with VP dialog classification, particularly in the no-hint condition ($p < .0001$). Hints significantly improved accuracy for VP dialogs ($p = .0026$), but did not significantly affect human dialog classification ($p = .2861$). Overall, when participants were provided with hints, overall classification accuracy increased from 55.0\% to 68.3\%, primarily due to improved VP detection (32.2\% → 51.7\%). However, nearly half of VP responses were still judged as human even with hints, suggesting a high degree of realism in VP-generated responses (see confusion matrix in \nameref{appendix:appendixb}, Table~\ref{tab:confusion}).

A second two-way ANOVA examined the effects of dialog type and hint condition on participants’ confidence ratings. There was a significant main effect of dialog type, $F(1, 90) = 4.04$, $p = .0475$, and a significant interaction between hint and dialog type, $F(1, 90) = 4.04$, $p = .0474$. However, the main effect of the hint condition alone was not significant, $F(1, 90) = 0.58$, $p = .4483$. Participants who received a hint reported greater confidence when judging human dialogs compared to VP dialogs ($p = .0055$; see Figure~\ref{fig:two_plots}, right), whereas no such pattern emerged in the no-hint condition ($p = .9754$). Between-group comparisons showed no significant difference in confidence for VP dialogs ($p = .3795$), but there was a marginal trend toward higher confidence in human dialog classification when a hint was provided ($p = .0531$).

\begin{table}[ht]
\centering
\caption{Log-Time Analysis by Dialog Type and Hint Condition.
SE = standard error; CI = confidence interval}
\label{tab:log_time}
\begin{tabular}{llccc}
\toprule
\textbf{Hint Condition} & \textbf{Dialog Type} & \textbf{Mean Log-Time} & \textbf{SE} & \textbf{95\% CI} \\
\midrule
With Hint     & VP-Generated & 3.13 & 0.164 & [2.80, 3.45] \\
Without Hint  & VP-Generated & 3.36 & 0.168 & [3.03, 3.69] \\
With Hint     & Human-Generated    & 3.05 & 0.164 & [2.72, 3.37] \\
Without Hint  & Human-Generated    & 3.23 & 0.168 & [2.89, 3.56] \\
\end{tabular}
\end{table}

Response times were log-transformed to reduce skewness and meet normality assumptions for parametric testing. As shown in Table~\ref{tab:log_time}, participants in the no-hint condition took slightly longer on average; however, differences across hint and dialog type conditions were not statistically significant (all $p > .05$). This suggests that the presence of a hint did not impact response latency, and that timing was unlikely to confound accuracy or confidence outcomes. Importantly, average classification accuracy for VP responses remained below chance level (<50\%), indicating that the VPs were often indistinguishable from real human responses—meeting the classical criterion for passing the Turing Test. Complete ANOVA results are reported in \nameref{appendix:appendixb} (Tables~\ref{tab:A1}–\ref{tab:A3}).

\subsection{Linguistic characteristics of VP and human simulated patient dialog turns}

\begin{figure}[!ht]
  \centering
  \includegraphics[width=0.73\textwidth]{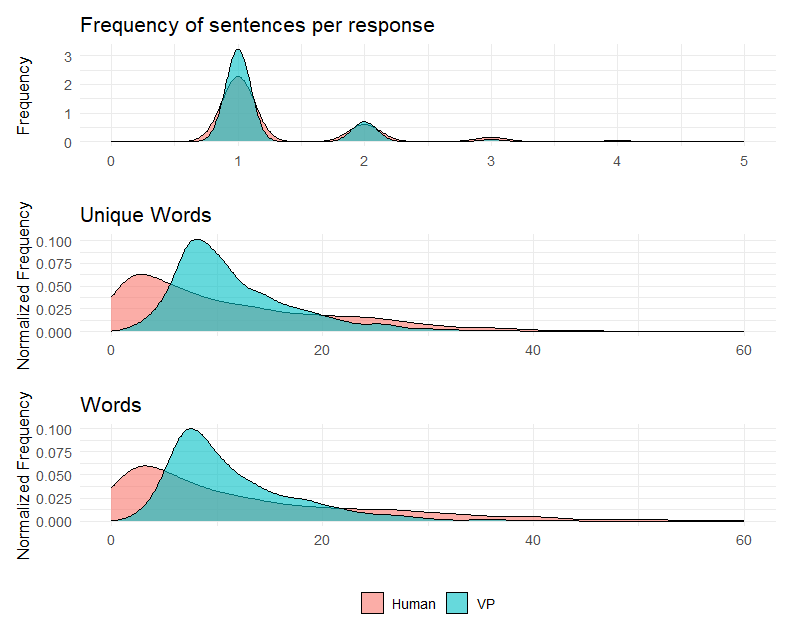} 
  \caption{Distributions of sentence count, unique words, and total word count per turn for human and VP responses.}
  \label{fig:three_plots}
\end{figure}

Across 44 conversations, we recorded a total of 2,194 conversation turns (question–answer pairs). Figure~\ref{fig:three_plots} presents the sentence length distributions for VP and human-simulated patient responses. While both groups typically responded with one sentence (median = 1), human responses were slightly longer on average (mean = 1.36) compared to VP responses (mean = 1.23). Notably, VPs tended to over-respond when human replies were brief, and conversely, under-responded when human responses were longer (see Figure~\ref{fig:responses_length} in \nameref{appendix:appendixb}).

To quantify distributional differences, we conducted two-sample Kolmogorov–Smirnov (KS) tests on total word count, sentence count, and number of unique words (Figure~\ref{fig:three_plots}). The largest divergences were observed in lexical variety ($D = 0.30$, $p < .001$) and total word count ($D = 0.29$, $p < .001$), indicating significant differences in verbosity and vocabulary use. Sentence count showed a smaller but still significant difference ($D = 0.06$, $p < .01$).

\begin{figure}[!ht]
  \centering
  \includegraphics[width=0.73\textwidth]{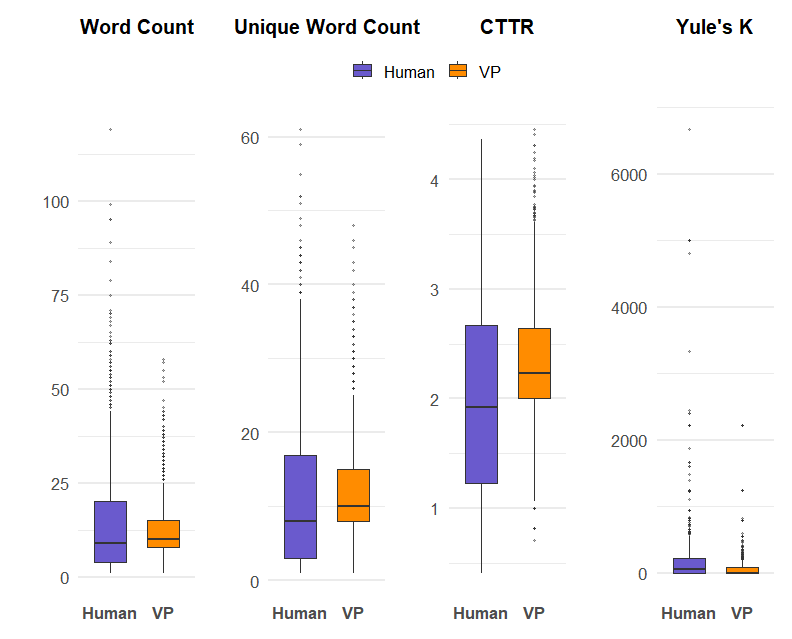} 
\caption{Lexical metric distributions computed per response for Human and VP answers. }  
\label{fig:lexical1}
\end{figure}

We further assessed lexical richness using multiple metrics. Interestingly, despite producing fewer words overall, VP responses exhibited higher lexical diversity across all measures  (Table~\ref{tab:lexical_metrics}). Lower Yule’s K scores for VPs indicate reduced lexical repetition, while higher MTLD, MATTR, CTTR, and HDD values suggest more varied and efficient vocabulary usage. To avoid aggregation bias, we also computed these metrics at the individual-response level (see Figure~\ref{fig:lexical1}). These findings suggest that while human responses were longer, they tended to be more repetitive, whereas VP responses were more compact yet lexically diverse.

\begin{table}[ht]
\centering
\begin{tabular}{lccccccc}
\toprule
           & Total Words & Unique Words & TTR   & RTTR & MTLD  & HDD   & MATTR \\
\midrule
Human      & 30,371      & 1,559        & 0.0513 & 8.95 & 40.33 & 0.798 & 0.817 \\
VP         & 27,517      & 1,660        & 0.0603 & 10.01 & 55.96 & 0.825 & 0.825 \\
\bottomrule
\end{tabular}
\caption{Lexical richness metrics comparing human and VP responses.}
\label{tab:lexical_metrics}
\end{table}


Part-of-speech (POS) frequency comparisons are shown in Table~\ref{tab:pos_tab} (also see Figure~\ref{fig:pos} in \nameref{appendix:appendixb}). While VP responses approximated human grammatical structure, Wilcoxon rank-sum tests revealed significant differences in several syntactic categories.Human responses contained more nouns and pronouns ($p < .001$), as well as a higher frequency of interjections ($p < .001$), such as “oh” or “hmm,” which convey emotion or conversational nuance. However, these differences may stem from the natural characteristics of real patient speech. Adverb usage was also higher in human responses ($p < .001$), whereas VPs used more adpositions ($p < .001$). A smaller but significant difference was found in particle usage, with humans employing them more frequently ($p = .045$).
By contrast, no significant differences were observed in the usage of verbs, auxiliaries, adjectives, conjunctions, or proper nouns ($p > .05$), suggesting that VPs closely mimic human behavior in core grammatical categories. However, the relative lack of expressive markers—such as interjections and adverbs—indicates a gap in pragmatic or emotional realism. Overall, these linguistic analyses provide quantitative insights into both the strengths and limitations of VP-generated language in clinical simulations.

\vspace{-0.25cm}

\begin{table}[ht]
\centering
\captionsetup{font=footnotesize}
\caption{
POS Tag Frequency Counts for Human and VP Responses. 
\textbf{ADJ} = adjective, \textbf{ADP} = adposition, \textbf{ADV} = adverb, \textbf{AUX} = auxiliary, 
\textbf{CCONJ} = coordinating conjunction, \textbf{DET} = determiner, \textbf{INTJ} = interjection, 
\textbf{NOUN} = noun, \textbf{PART} = particle, \textbf{PRON} = pronoun, \textbf{PROPN} = proper noun, 
\textbf{SCONJ} = subordinating conjunction, \textbf{VERB} = verb.
}
\label{tab:pos_tab}
\resizebox{\textwidth}{!}{%
\begin{tabular}{lrrrrrrrrrrrrr}
\toprule
\textbf{Group} & \textbf{ADJ} & \textbf{ADP} & \textbf{ADV} & \textbf{AUX} & \textbf{CCONJ} & \textbf{DET} & \textbf{INTJ} & \textbf{NOUN} & \textbf{PART} & \textbf{PRON} & \textbf{PROPN} & \textbf{SCONJ} & \textbf{VERB} \\
\midrule
Human & 1674 & 2190 & 2982 & 1899 & 1082 & 2073 & 2651 & 5878 & 496 & 5460 & 213 & 663 & 3451 \\
VP    & 1915 & 2296 & 2481 & 1909 & 1056 & 1995 & 2126 & 7020 & 370 & 4044 & 181 & 346 & 3188 \\
\bottomrule
\end{tabular}%
}
\end{table}

\vspace{-0.25cm}
\subsection{Semantic similarity analysis}
We assessed semantic similarity between VP and human-simulated patient responses at two levels using cosine similarity of BioClinicalBERT embeddings:

\begin{itemize}
    \item \textbf{Response-level similarity:} For each transcript, we computed cosine similarity scores between VP and human responses at the turn level. These pairwise similarities were then averaged per transcript. The mean response-level similarity across all transcripts was 0.871 (SD = 0.1301), indicating a high degree of semantic overlap at the local interaction level.
    
    \item \textbf{Whole-transcript similarity:} To evaluate global semantic alignment, all patient responses within each transcript were concatenated into a single block of text. We then computed cosine similarity between the aggregated VP and human embeddings. This analysis yielded a mean whole-transcript similarity of 0.842 (SD = 0.045), reflecting consistent semantic correspondence across full dialogues.
\end{itemize}

\vspace{-0.5cm}
\section{Discussion} 
\vspace{-0.25cm}

In this study, We developed 44 virtual patients (VPs) using GPT-4o, prompted with structured illness scripts based on doctor–simulated patient dialogues across 17 categories \cite{fareez2022dataset}. To evaluate realism, we combined expert review, linguistic and semantic analysis, and a Turing Test. Experts rated responses for hallucinations, omissions, repetitions, and completeness, finding hallucinations most common (2.8\%) but overall strong interrater reliability and high response quality.

In the Turing Test, those given a hint about VP traits identified VPs more accurately and showed higher confidence when judging human responses—but not VPs. Still, response times were unaffected by hints or response type. Human responses were easier to recognize, but VP replies often mimicked human language well enough to pass as real, highlighting the realism of LLM-generated interactions.

Linguistic analysis of 2,000+ turns showed that both VPs and humans used similar grammatical patterns, though human responses were slightly longer and more expressive, while VPs were more concise and lexically varied. Semantic analysis using BioClinicalBERT confirmed that VP dialogues preserved key clinical information and closely resembled real patient responses, without requiring expert review.

Compared to prior studies \cite{cook2024creating, holderried2024generative, munoz2024contrasting}, our work stands out by using structured illness scripts and combining expert and computational evaluations. Unlike earlier efforts with limited cases and inconsistent prompting, we present a scalable framework applied across 44 VP scenarios.

This study has several limitations. First, we used only GPT-4o, which may limit generalizability. The illness scripts were based on simulated doctor–patient interactions, and doctor questions were pre-scripted, reducing the adaptability of the dialogue. The VPs focused solely on clinical symptoms, without incorporating behavioral aspects or demographic diversity. We used a zero-shot prompting approach; exploring few-shot prompts or fine-tuning could enhance realism. Finally, all cases were restricted to respiratory conditions, limiting the diversity across clinical specialties.


\bibliography{references}

\newpage

\appendix
\noindent
\begin{minipage}{\textwidth}
\vspace*{-3em} 

\section*{Appendix A}
The code is available at \href{https://github.com/Reyhanehrhp7/Virtual-Patients-Using-Open-AI-API-2024}{https://github.com/Reyhanehrhp7/Virtual-Patients-Using-Open-AI-API-2024}. 

\label{appendix:appendixa}

\begin{tcolorbox}[
  colback=yellow!10,
  colframe=orange!80!black,
  colbacktitle=orange!80!black,
  coltitle=white,
  title=Box \refstepcounter{mybox}\label{box:example_script}\themybox: Excerpt from an Example Ilness Script,
  fonttitle=\bfseries,
  boxrule=0.4pt,
  sharp corners=south,
  before skip=6pt, after skip=6pt,
  left=4pt, right=4pt, top=4pt, bottom=4pt,
]
\footnotesize
\raggedright
\textbf{You are going to play the role of a medical patient.}

\vspace{0.8em}
\textbf{The patient details are as follows:}

\textbf{Name:} Unknown to the transcript. \\
\textbf{Age:} Unknown to the transcript.\\
\textbf{Chief Complaint:} Cough that brings up gunk. \\
\textbf{Symptoms:} Green, sometimes yellow, gunk; couple Kleenexes worth of sputum. \\
\textbf{Duration:} Past few years. \\
\textbf{Nature and rating of pain:} Coughing is always pretty bad, worse with exertion like walking up stairs. \\
\textbf{Pain/symptom progression:} Cough is getting worse now, worse in the morning. Deep breaths make it worse. \\
\textbf{Additional symptoms:} Low energy, feeling tired, slightly swollen, possible weight gain, trouble breathing with exertion, frequent urination at night, poor sleep. \\
\textbf{Medical History:} High blood pressure, C-section for birth of daughter, hospitalized for a few days after that. \\
\textbf{Medications:} Lisinopril. \\
\textbf{Immunizations:} Unknown to the transcript. \\
\textbf{Family History:} Father had a heart attack at age 78. \\
\textbf{Alcohol, tobacco, marijuana, other drugs:} No recreational drugs; three to four glasses of wine per week (5–6 oz per glass); smokes cigarettes (1–2 packs/day for the last 40 years). \\
\textbf{Living Conditions:} Lives in an apartment with husband and daughter. \\
\textbf{Recent Travel:} No recent travel. \\
\textbf{Occupation:} Worker at local grocery store. \\
\textbf{Other:} Uses one pillow. Lost five pounds in the last few months. Recently tested negative for tuberculosis. Has a decent diet but does not exercise. \\

\textbf{Answer with the main complaint initially. Respond with other symptoms only when asked or when relevant in the conversation, maintain conversational language.\\
If a response slot says "unknown to the patient", respond in a way that indicates the patient doesn't know and has no information about it.\\
If a response slot says "the patient is unsure", respond in a way that indicates uncertainty and lack of surety.\\
If a response slot says "absent in the patient", respond with a response indicating such.\\
If a response slot says "unknown to the transcript", make up a relevant response.}
\end{tcolorbox}

\begin{figure}[H]
\centering
\includegraphics[width=\textwidth]{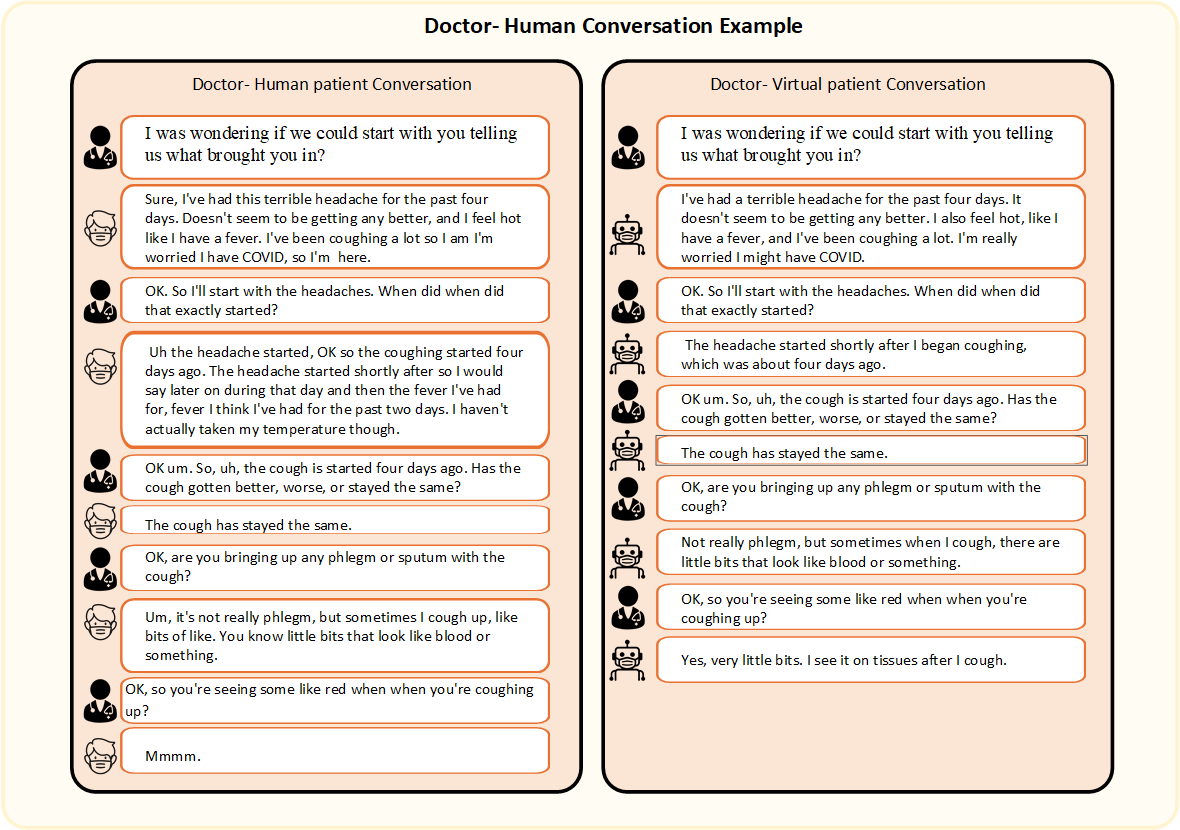}
\caption{Figure~\ref{fig:conversation} shows a side-by-side example comparing a doctor--human conversation and a doctor--VP interaction. Both interactions follow the same clinical prompt and use the same \textbf{doctor} utterances.}
\label{fig:conversation}
\end{figure}
\end{minipage}

\begin{tcolorbox}[
  colback=yellow!10,
  colframe=orange!80!black,
  colbacktitle=orange!80!black,
  coltitle=white,
  title=Box \refstepcounter{mybox}\label{box:complete_example_script}\themybox: A complete example of Doctor-VP Interaction,
  fonttitle=\bfseries,
  boxrule=0.4pt,
  sharp corners=south,
  before skip=6pt, after skip=6pt,
  left=4pt, right=4pt, top=4pt, bottom=4pt,
]
\footnotesize
\raggedright
\textbf{User}: I was wondering if you could tell me what brought you in.\\
\textbf{ChatGPT}: I've been having a lot of trouble with coughing and shortness of breath lately. The cough has been there for the past five years, but it's gotten a lot worse in the last three days.\\

\textbf{User}: OK, when did the cough start?\\
\textbf{ChatGPT}: The cough started about five years ago.\\

\textbf{User}: OK, did anything happen 3 days ago that has brought the cough on?\\
\textbf{ChatGPT}: No, nothing specific happened three days ago. The cough just seemed to get a lot worse all of a sudden.\\

\textbf{User}: OK, is there any particular time of day that the cough is worse, like morning, night?\\
\textbf{ChatGPT}: Yes, the cough is definitely worse in the morning.

\textbf{User}: OK, and are you bringing up any sputum?\\
\textbf{ChatGPT}: Yes, I'm bringing up sputum. It used to be clear, but now it's yellow and sometimes green. And I'm bringing up more sputum than before.

\textbf{User}: OK. With your cough prior to three days ago were you bringing up sputum with that?\\
\textbf{ChatGPT}: Yes, I was bringing up sputum before, but it was clear back then and there was less of it compared to now.

\textbf{User}: OK, and are you bringing up more sputum now, the same or less?\\
\textbf{ChatGPT}: I'm bringing up more sputum now than before.

\textbf{User}: OK, have you noticed any blood?\\
\textbf{ChatGPT}: No, I haven't noticed any blood in the sputum.

\textbf{User}: OK, and you mentioned you've been you've had a wheeze, when you do you notice that?\\
\textbf{ChatGPT}: I mostly notice the wheeze at night.

\textbf{User}: OK, um and what types of activities have you had shortness of breath?\\
\textbf{ChatGPT}: I've been getting short of breath just from walking and doing day-to-day things.

\textbf{User}: OK. And have you been experiencing any other symptoms besides the cough, wheeze and shortness of breath?\\
\textbf{ChatGPT}: Yes, I've also been feeling fatigued, though I sleep normally. Sometimes I get a dull chest pain from coughing so frequently.

\textbf{User}: OK. So a bit more tired. How has your sleep been?\\
\textbf{ChatGPT}: My sleep has been normal, despite feeling more fatigued during the day.

\textbf{User}: OK, have you had any ear pain or eye pain?\\
\textbf{ChatGPT}: No, I haven’t had any ear pain or eye pain.

\textbf{User}: Any discharge from either, either your ears or eyes?\\
\textbf{ChatGPT}: No, I haven't had any discharge from my ears or eyes.

\textbf{User}: OK. Have you had a runny nose?\\
\textbf{ChatGPT}: No, I haven’t had a runny nose.

\textbf{User}: Or felt congested?\\
\textbf{ChatGPT}: No, I haven't felt congested either.

\textbf{User}: OK, have you had a sore throat?\\
\textbf{ChatGPT}: No, I haven't had a sore throat.

\textbf{User}: Or any difficulty swallowing?\\
\textbf{ChatGPT}: No, I haven't had any difficulty swallowing either.

\textbf{User}: Changes to your sense of smell or taste?\\
\textbf{ChatGPT}: No, I haven't noticed any changes to my sense of smell or taste.

\textbf{User}: OK, have you had any changes in appetite?\\
\textbf{ChatGPT}: No, my appetite has been normal.

\textbf{User}: Alright, and have you had any weight loss over the last several months?\\
\textbf{ChatGPT}: No, I haven't had any weight loss over the last several months.

\textbf{User}: OK. How about have you had any fevers or chills?\\
\textbf{ChatGPT}: No, I haven't had any fevers or chills.

\textbf{User}: Nights, how about night sweats?\\
\textbf{ChatGPT}: No, I haven't had any night sweats.

\textbf{User}: Have you had any chest pain?\\
\textbf{ChatGPT}: Yes, I've had some dull chest pain, but it seems to be from coughing frequently.

\textbf{User}: OK, do you get chest pain when you exert yourself physically?\\
\textbf{ChatGPT}: No, the chest pain doesn't seem to get worse with physical exertion; it mostly happens because of the frequent coughing.

\textbf{User}: OK, so it's just after the coughing?\\
\textbf{ChatGPT}: Yes, the chest pain is mostly just from coughing so much.

\textbf{User}: OK and how would you describe the character of the chest pain? Some people will say sharp, dull, achy.\\
\textbf{ChatGPT}: I'd describe the chest pain as dull.

\end{tcolorbox}

\begin{tcolorbox}[
  colback=yellow!10,
  colframe=orange!80!black,
  colbacktitle=orange!80!black,
  coltitle=white,
  title=Box \refstepcounter{mybox}\label{box:conversation2}\themybox: Conversation Cont.,
  fonttitle=\bfseries,
  boxrule=0.4pt,
  sharp corners=south,
  before skip=6pt, after skip=6pt,
  left=4pt, right=4pt, top=4pt, bottom=4pt,
]
\footnotesize
\raggedright

\textbf{User}: OK, and have you had any like heart palpitations, feeling like your heart's racing?\\
\textbf{ChatGPT}: No, I haven't experienced any heart palpitations or felt like my heart is racing.

\textbf{User}: And how about any episodes where you felt like you were going to faint or, yeah, pass out?\\
\textbf{ChatGPT}: No, I haven't had any episodes where I felt like I was going to faint or pass out.
\textbf{User}: OK. Uhm and have you had any swelling in your feet, ankles, hands?\\
\textbf{ChatGPT}: No, I haven't noticed any swelling in my feet, ankles, or hands.

\textbf{User}: OK. And then have you had any urinary problems?\\
\textbf{ChatGPT}: No, I haven't had any urinary problems.

\textbf{User}: Any bowel problems?\\
\textbf{ChatGPT}: No, I haven't had any bowel problems either.

\textbf{User}: OK, and then any joint pains or muscle aches?\\
\textbf{ChatGPT}: No, I haven't had any joint pains or muscle aches.

\textbf{User}: And how about any rashes anywhere?\\
\textbf{ChatGPT}: No, I haven't had any rashes.

\textbf{User}: OK, and then, sorry, have you had any nausea or vomiting?\\
\textbf{ChatGPT}: No, I haven't had any nausea or vomiting.

\textbf{User}: OK. In terms of your past medical history, have you been diagnosed with any conditions before?\\
\textbf{ChatGPT}: Yes, I have been diagnosed with high blood pressure and diabetes.

\textbf{User}: I see. Are you on any medications for those?\\
\textbf{ChatGPT}: Yes, I'm taking a beta blocker for my high blood pressure and metformin for my diabetes. I also take calcium and vitamin D supplements.

\textbf{User}: OK, uh, do you, are you taking any other medications?\\
\textbf{ChatGPT}: No, just the ones I mentioned: the beta blocker, metformin, calcium, and vitamin D.

\textbf{User}: Any surgeries in the past?\\
\textbf{ChatGPT}: No, I haven't had any surgeries in the past.

\textbf{User}: Any allergies to medications?\\
\textbf{ChatGPT}: No, I don't have any allergies to medications.

\textbf{User}: OK. Are your immunizations up to date?\\
\textbf{ChatGPT}: Yes, my immunizations are up to date, but I haven't gotten the pneumococcal vaccine.

\textbf{User}: Yeah, you're you're right. So for um, in Canada or anybody or in Ontario, at least, anybody who's older than 65, it's recommended to get this vaccine to protect against certain types of bacterial infections type which can cause pneumonia. So it would be good something that you can set up with your family doctor.\\
\textbf{ChatGPT}: That sounds like a good idea. I'll make sure to talk to my family doctor about getting the pneumococcal vaccine.

\textbf{User}: Yeah, and so are you, so do you smoke cigarettes?\\
\textbf{ChatGPT}: I used to smoke a pack a day for 45 years, but I stopped two years ago.

\textbf{User}: OK, and how much do you smoke or how much did you use to smoke?\\
\textbf{ChatGPT}: I used to smoke a pack a day.

\textbf{User}: OK, that's great that you were able to stop. Do you drink alcohol?\\
\textbf{ChatGPT}: Yes, I drink alcohol occasionally, about every two weeks.

\textbf{User}: OK, and any triggers that you can think of you were exposed to either at work or home like chemicals or things that you were exposed to that you had to breathe, that you breathe in?\\
\textbf{ChatGPT}: No, I haven't been exposed to any triggers like chemicals at work or home.

\textbf{User}: OK, have you traveled anywhere?\\
\textbf{ChatGPT}: No, I haven't traveled anywhere recently.

\textbf{User}: Alright, and have you been near anybody who's been sick or has similar symptoms?\\
\textbf{ChatGPT}: No, I haven't been near anyone who is sick or has similar symptoms.

\textbf{User}: Alright, and anybody in the family have a heart or lung condition?\\
\textbf{ChatGPT}: No, but there is liver cancer in my family.

\textbf{User}: Any cancers in the family?\\
\textbf{ChatGPT}: Yes, there's liver cancer in my family.

\textbf{User}: OK, and, I think that.. OK, and that was all I wanted to ask on history. Was there anything else that you wanted to tell me today?\\
\textbf{ChatGPT}: No, I think we covered everything. Thanks for asking.

\textbf{User}: Yeah, certainly so umm, it, it sounds like it could be maybe a viral infection that I brought on this cough and worsen the sputum. This could be a bacterial infection, it could be a COPD exacerbation and also one of the other things that we would have to consider right now is Covid as well and so we will do a PCR nasal swab for Covid and a chest Xx-ray and some other investigations as well and uhm, we will try to get to the bottom of what's causing your your shortness of breath.\\
\textbf{ChatGPT}: That sounds good. Thank you for explaining. I appreciate you looking into this and helping me figure out what's going on.
\end{tcolorbox}

\newpage

\section*{Appendix B}
\label{appendix:appendixb}

\begin{table}[ht]
\centering
\caption{Estimated Marginal Means and Pairwise Comparisons of Confidence Ratings by Hint Condition and Dialog Type}
\label{tab:A1}
\begin{tabular}{lllll}
\toprule
\textbf{Hint Condition} & \textbf{Dialog Type} & \textbf{Mean Confidence} & \textbf{SE} & \textbf{95\% CI} \\
\midrule
With Hint    & VP-Generated & 2.63 & 0.0974 & [2.44, 2.82] \\
Without Hint & VP-Generated & 2.75 & 0.0995 & [2.55, 2.95] \\
With Hint    & Human-Generated    & 3.02 & 0.0974 & [2.83, 3.21] \\
Without Hint & Human-Generated    & 2.75 & 0.0995 & [2.55, 2.95] \\
\bottomrule
\end{tabular}
\end{table}

\begin{table}[ht]
\centering
\caption{Confidence Ratings by Human Evaluators in the Turing Test}
\label{tab:A2}
\begin{tabular}{lllll}
\toprule
\textbf{Comparison} & \textbf{Mean Difference} & \textbf{SE} & \textbf{$t$(90)} & \textbf{\textit{p}-value} \\
\midrule
With Hint: Human vs. VP   & -0.39 & 0.138 & -2.84 & .0055 * \\
Without Hint: Human vs. VP &  0.004 & 0.141 & 0.03 & .9754 \\
VP: With vs. Without Hint  & -0.12 & 0.139 & -0.88 & .3795 \\
Human: With vs. Without Hint     &  0.27 & 0.139 & 1.96 & .0531 † \\
\bottomrule
\end{tabular}

\vspace{1ex}
\raggedright
\textbf{Note.} CI = Confidence Interval. ** $p < .01$. † Trend-level difference ($p < .10$).
\end{table}

\begin{table}[ht]
\centering
\caption{Summary of ANOVA Results for Correct Responses, Confidence Level, and Log-Transformed Response Time. 
\label{tab:A3}
Significance levels: *: $p$ $<$ 0.05, **: $p$ $<$ 0.01, ***: $p$ $<$ 0.001.}
\label{tab:anova_summary}
\begin{tabular}{llrrrrl}
\toprule
\textbf{Variable} & \textbf{Source} & \textbf{Sum Sq} & \textbf{Df} & \textbf{Mean Sq} & \textbf{F} & \textbf{$p$-value} \\
\midrule
\multirow{4}{*}{Correct Responses} 
    & Group               & 40.50   & 1  & 40.50   & 8.689   & 0.0041 ** \\
    & Transcripts         & 360.20  & 1  & 360.20  & 77.34   &  0.001 *** \\
    & Group $\times$ Transcripts & 9.50 & 1 & 9.50 & 2.045 & 0.1562\\
    & Residuals           & 419.10  & 90 & 4.66    &         &        \\
\midrule
\multirow{4}{*}{Confidence Level} 
    & Group               & 0.132   & 1  & 0.132   & 0.580   & 0.4483    \\
    & Transcripts         & 0.920   & 1  & 0.920   & 4.039   & 0.0475 *   \\
    & Group $\times$ Transcripts & 0.921 & 1 & 0.921 & 4.042 & 0.0474 * \\
    & Residuals           & 20.50   & 90 & 0.228   &         &           \\
\midrule
\multirow{4}{*}{Log Time} 
    & Group               & 1.00    & 1  & 1.00    & 1.549   & 0.217      \\
    & Transcripts         & 0.27    & 1  & 0.27    & 0.412   & 0.523       \\
    & Group $\times$ Transcripts & 0.01 & 1 & 0.01 & 0.022 & 0.881  \\
    & Residuals           & 58.24   & 90 & 0.647   &         &        \\
\bottomrule
\end{tabular}
\end{table}

\begin{table}[ht]
\centering
\caption{Turing Test Confusion Matrices With and Without Hint}
\label{tab:confusion}
\begin{minipage}{0.48\textwidth}
\centering
\textbf{(a) With Hint}
\vspace{0.2cm}

\begin{tabular}{l|cc|c}
                & Human & VP & Total \\
\hline
Judged Human & 203 & 116 & 319 \\
Judged VP   & 37  & 124 & 161 \\
\hline
Total  & 240 & 240 & 480 \\
\end{tabular}
\end{minipage}
\hfill
\begin{minipage}{0.48\textwidth}
\centering
\textbf{(b) Without Hint}
\vspace{0.2cm}

\begin{tabular}{l|cc|c}
                & Human & VP & Total \\
\hline
Judged Human & 179 & 156 & 335 \\
Judged VP   & 51  & 74  & 125 \\
\hline
Total   & 230 & 230 & 460 \\
\end{tabular}
\end{minipage}
\end{table}

\begin{figure}[!ht]
  \centering
  \includegraphics[width=0.8\textwidth]{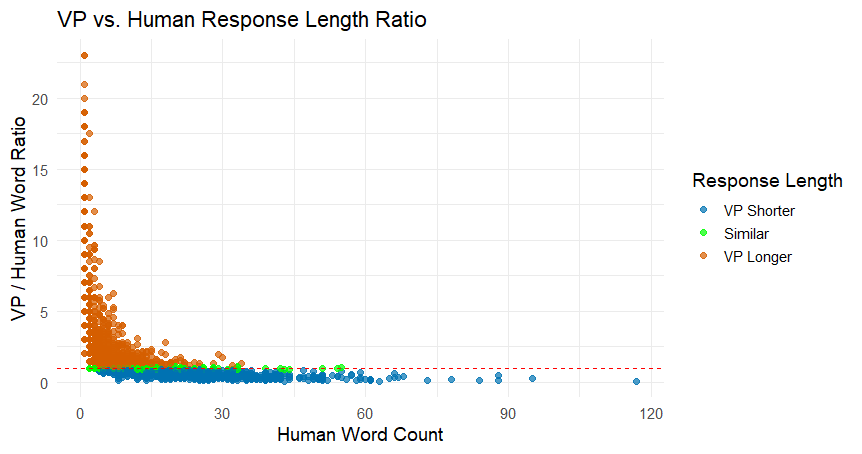} 
\caption{VP vs. Human Response length Ration}  
\label{fig:responses_length}
\end{figure}

\begin{figure}[!ht]
  \centering
  \includegraphics[width=0.75\textwidth]{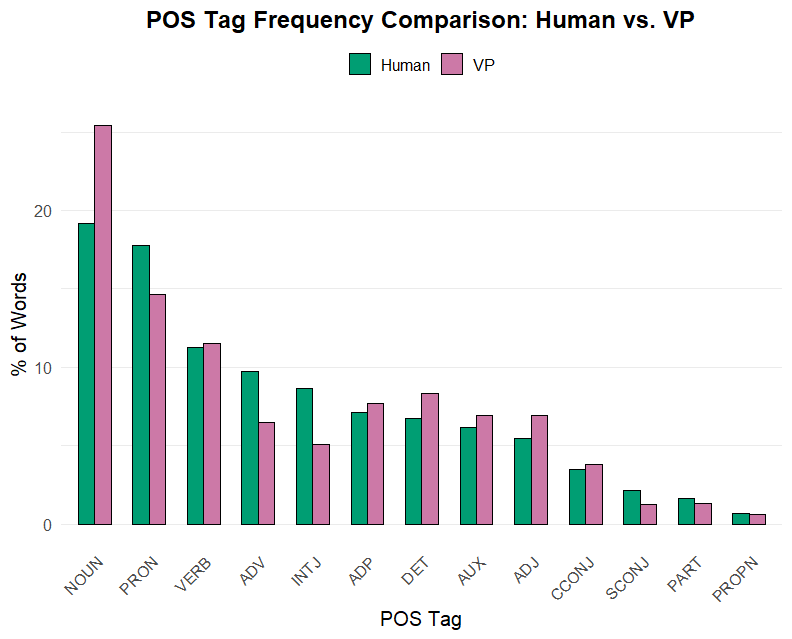} 
\caption{Percentage differences in the use of POS category for VP responses in comparison to human simulated patients. }  
\label{fig:pos}
\end{figure}


\newpage
\section*{Appendix C}
\label{appendix:appendixc}

\subsection*{Turns with Omissions}
This metric refers to the number of conversations where ChatGPT-4o omits important information from the illness script that is necessary for making an accurate diagnosis.

\textbf{Illness Script:}
\begin{quote}
The patient has no family history of lung conditions. Their mother had a heart attack at the age of 56. The patient has not traveled recently.
\end{quote}

\textbf{Example of Conversation:}
\begin{quote}
D: You are OK. Any family history of lung conditions? \\
P: No lung condition. \\
D: But how about heart? \\
P: Yeah, it's just my mom had a heart attack when she was 56.
\end{quote}

\textbf{Example of a Turn with an Omission:}
\begin{quote}
D: You are OK. Any family history of lung conditions? \\
P: No lung condition. \\
D: But how about heart? \\
P: No heart issues.
\end{quote}

\subsection*{Turns with Hallucinations (Could Influence Diagnosis)}
Hallucinations in this context mean instances where ChatGPT-4o provides false, fabricated, or misleading information relative to the illness script, which could lead to incorrect diagnoses or inappropriate treatment plans.

\textbf{Illness Script:}
\begin{quote}
The patient has no family history of lung conditions. Their mother had a heart attack at the age of 56. The patient has not traveled recently.
\end{quote}

\textbf{Example of Conversation:}
\begin{quote}
D: You are OK. Any family history of lung conditions? \\
P: No lung condition. \\
D: But how about heart? \\
P: Yeah, it's just my mom had a heart attack when she was 56.
\end{quote}

\textbf{Example of a Turn with a Hallucination:}
\begin{quote}
D: But how about heart? \\
P: Yeah, it's my mom and dad had a heart attack.
\end{quote}

\subsection*{Turns with Inappropriate Repetitions}
This refers to the number of conversation turns where ChatGPT-4o repeats information or questions inappropriately in a way that is not human-like.

\textbf{Example of a Turn with a Repetition:}
\begin{quote}
D: Can you describe the nature of your pain? \\
P: The pain is sharp and localized to my lower abdomen. It started suddenly two days ago. \\
D: Have you experienced any nausea or vomiting? \\
P: Yes, I've had nausea and vomiting. Additionally, the pain is still sharp and localized to my lower abdomen, and it hasn't changed in intensity since it started suddenly two days ago.
\end{quote}

\bibliographystyle{plainnat}

\end{document}